\documentclass[
    groupedaddress,
    superscriptaddress,
    preprint,
     amsmath,amssymb,
     aps,
     prl,
     showpacs,
    floatfix,
    ]{revtex4-2}
 \usepackage{comment}
\usepackage{graphicx}
\usepackage{dcolumn}
\usepackage{bm}
\usepackage{dirtytalk}

\usepackage{hyperref}
\hypersetup{
    colorlinks=true,
    linkcolor=blue,
    citecolor=blue,
    urlcolor=blue
    }
\renewcommand{\doi}[1]{\href{https://doi.org/#1}{\nolinkurl{#1}}}
\newcommand{\ZrTe}{ZrTe$_2$}
    
\begin{document}
\title{Emergent superconductivity and non-reciprocal transport in a van der Waals Dirac semimetal/antiferromagnet heterostructure}

\author{Saurav Islam}
\email{ski5160@psu.edu}
\author{Max Stanley}
\author{Anthony Richardella}
\affiliation{Department of Physics, Pennsylvania State University, University Park, Pennsylvania 16802, USA}
\affiliation{Materials Research Institute, Pennsylvania State University, University Park, Pennsylvania 16802, USA}
\author{Seungjun Lee}
\affiliation{Department of Electrical and Computer Engineering, University of Minnesota, 
Minneapolis, MN 55455, USA.}
\author{Kalana D. Halanayake}
\affiliation{Department of Chemistry, Pennsylvania State University, University Park, Pennsylvania 16802, USA}
\author{Sandra Santhosh}
\affiliation{Department of Physics, Pennsylvania State University, University Park, Pennsylvania 16802, USA}
\author{Danielle Reifsnyder Hickey}
\affiliation{Department of Chemistry, Pennsylvania State University, University Park, Pennsylvania 16802, USA}
\affiliation{Department of Materials Science and Engineering, Pennsylvania State University, University Park, Pennsylvania 16802, USA}
\affiliation{Materials Research Institute, Pennsylvania State University, University Park, Pennsylvania 16802, USA}
\author{Tony Low}
\affiliation{Department of Electrical and Computer Engineering, University of Minnesota, Minneapolis, MN, 55455, USA}
\affiliation{School of Physics and Astronomy, University of Minnesota, Minneapolis, MN, 55455, USA}
\author{Nitin Samarth}
\email{nxs16@psu.edu}
\affiliation{Department of Physics, Pennsylvania State University, University Park, Pennsylvania 16802, USA}
\affiliation{Department of Materials Science and Engineering, Pennsylvania State University, University Park, Pennsylvania 16802, USA}
\affiliation{Materials Research Institute, Pennsylvania State University, University Park, Pennsylvania 16802, USA}
    
\begin{abstract}
    
We investigate emergent superconductivity and non-reciprocal transport (magnetochiral anisotropy, superconducting diode effect) at the heterointerface of two non-superconducting van der Waals (vdW) materials, the Dirac semimetal ZrTe$_2$ and the antiferromagnetic iron chalcogenide FeTe, grown using molecular beam epitaxy. We show from electrical transport measurements that two-dimensional (2D) superconductivity arises at the heterointerface below a critical temperature $T_c \sim 10$~K. In the superconducting transition region, non-reciprocal transport, characterized by the magneto-chiral anisotropy, exhibits a magnitude comparable to that observed in topological insulators, and is enhanced by a factor of three when the heterostructure is capped with a 2D vdW ferromagnet (CrTe$_2$). Below $T_c$, the superconducting diode effect exhibits an efficiency of 29\%. With strong spin-orbit coupling in ZrTe$_2$, these epitaxial heterostructures provide an attractive epitaxial vdW platform for exploring unconventional superconductivity in Dirac semimetals and for developing non-reciprocal devices for superconducting electronics.
    
\end{abstract}
\maketitle

    
Topological materials interfaced with superconductivity have attracted significant interest as potential platforms for quantum applications, starting with proposals involving topological insulators~\cite{fu2008superconducting,qi2011topological} and later topological semimetals~\cite{bednik2015superconductivity,kobayashi2015topological}. Dirac semimetals (DSMs)  ~\cite{armitage2018weyl,burkov2016topological,borisenko2014experimental,yan2017topological,ong2021experimental,xiong2015evidence,shekhar2015extremely} are particularly appealing within this context for two reasons. First, the non-trivial topology of the bands in the normal state, comprised of Dirac points and surface Fermi loops, is predicted to result in bulk point nodes and surface Majorana modes in the superconducting state, potentially leading to unconventional pairing in some materials~\cite{kobayashi2015topological}. Second, combining a DSM simultaneously with superconductivity and ferromagnetism could provide a platform for realizing monopole superconductivity~\cite{li2018topological,bobrow2020monopole,sun2019vortices}. Finally, such hybrid magnetic/topological/superconductor heterostructures that combine broken inversion and time-reversal symmetry with strong spin-orbit coupling provide the necessary ingredients for designing high efficiency non-reciprocal devices (such as field-free Josephson diodes) for superconducting electronics \cite{Davydova_SciAd,Wu_NatMat}. 
    
Experimental studies of inducing superconductivity in DSMs have largely centered on the canonical material, Cd$_3$As$_2$, using mesoscopic point contacts \cite{aggarwal2016unconventional,wang2016observation} or pressure \cite{he2016pressure} in bulk crystals, and the proximity effect in nanoplates ($\approx100$~nm)~\cite{huang2019proximity,chu2023broad,li2020fermi} and thin films ~\cite{suslov2019observation,rashidi2023induced,rashidi2024vortex}. Despite these concerted efforts, a materials platform that enables the induction of superconductivity in a well-controlled heterostructure geometry has remained elusive, motivating the continued exploration of alternate strategies to introduce superconductivity in a DSM. 


Heterostructures that interface FeTe with other Te-based materials provide an attractive opportunity in this context~\cite{he2014two,Yasuda_NComm,liang2020studies,yao2021hybrid,yao2022superconducting,tkavc2023multiphase,yao2024mystery,yi2023dirac,hemianCBST,yuan2024coexistence,Yan_2025}. Until very recently, the prevailing view considered FeTe as an antiferromagnet (AFM) in its stoichiometric phase, making a transition to a superconducting phase only for non-stoichiometric conditions or under pressure; however, defect-free FeTe has recently been shown to be {\it intrinsically superconducting} \cite{yan2026stoichiometric}. Here, we realize emergent 2D superconductivity and non-reciprocal transport (including an efficient superconducting diode effect) by combining FeTe with  the van der Waals (vdW) DSM ZrTe$_2$ \cite{tsipas2018massless} and with a hybrid vdW DSM/FM heterostructure (ZrTe$_2$/CrTe$_2$) \cite{ou2022zrte2}. 
    
We demonstrate the emergence of superconductivity and non-reciprocal transport in these hybrids via magneto-resistance measurements in the linear and non-linear response regimes as a function of temperature, magnetic field, and current density. Analysis of the current-voltage characteristic (IVC) reveals the emergence of two-dimensional (2D) superconductivity below $T_c \sim 10$~K. Second harmonic electrical transport measurements with an in-plane magnetic field demonstrate non-reciprocal transport, characterized by a magneto-chiral anisotropy coefficient of similar magnitude to that found in topological insulator/FeTe heterostructures~\cite{Yasuda_NComm}. The latter study attributed the pronounced non-reciprocal transport to the spin-momentum correlation in helical Dirac surface states; our observation of similar behavior in a DSM with spin-degenerate bulk Dirac bands shows that this is not essential. When the FeTe/ZrTe$_2$ heterostructure is capped with a 2D vdW ferromagnet (CrTe$_2$), this figure-of-merit is remarkably enhanced threefold, akin to the behavior found in heterostructures that directly interface CrTe$_2$ with FeTe \cite{Yan_2025}, although the 2D FM here is separated from the FeTe interface by a several nm thick DSM. In particular, when these hybrids are fully superconducting, we demonstrate a superconducting diode effect with efficiency as high as $29$~\%. Our results demonstrate a new platform that allows systematic manipulation of the interplay of topology, superconductivity, and magnetism at conveniently accessible temperatures ($T \geq 4.2$~K).
    
ZrTe$_2$ has a layered crystal structure with Zr atoms sandwiched between Te atoms that are stacked along the c-axis (Fig.~\ref{1}(a))~\cite{tsipas2018massless,ou2022zrte2}. We refer to ZrTe$_2$ film thickness in terms of the unit cell (UC) thickness along c-axis.  Each unit cell of FeTe is composed of one layer of Fe atoms sandwiched by two layers of Te atoms (Fig.~\ref{1}(b))~\cite{subedi2008density,chen2009electronic}. Prior neutron diffraction studies of bulk crystals of FeTe indicate that the spins of the Fe atoms align diagonally across the Fe–Fe square lattice, giving rise to a bicollinear AFM, accompanied by a tetragonal to a monoclinic structural phase transition. The FeTe/ZrTe$_2$ heterostructures are grown on SrTiO$_3$ (100) substrates by molecular beam epitaxy (MBE) (See Supplemental Material (SM)~\cite{Islam_Supp}).
    
The reflection high-energy electron diffraction (RHEED) shows streaky patterns consistent with two-dimensional growth and the formation of twinned domains (Fig.~\ref{1}(c))~\cite{ichimiya2004reflection,hasegawa2012reflection}. Detailed microscopic structural information is obtained using high-angle annular dark-field scanning transmission electron microscopy (HAADF-STEM) imaging of cross-sectional samples: this shows a relatively sharp interface between FeTe and ZrTe$_2$ (Fig.~\ref{1}(d)) (See SM~\cite{Islam_Supp}). EDX mapping of the cross-section reveals the expected composition (Fig.~\ref{1}(e)). The EDX signal of Zr indicates that ZrTe$_2$ layers are located on the FeTe layers. Additionally, signal from oxygen is also detected in the EDX signal near the ZrTe$_2$ layer, resulting from some oxidation of ZrTe$_2$. (Fig.~\ref{1}(f)). We characterize the crystalline quality of the heterostructures {\it ex situ} using X-ray diffraction (XRD) $2\theta-\omega$ scans (Fig.~\ref{1}(g)) and observe the expected peaks corresponding to SrTiO$_3$(100), ZrTe$_2$, and FeTe layers (See SM~\cite{Islam_Supp}).  XRD $\phi$ scans demonstrate that FeTe grows aligned to the SrTiO$_3$ substrate while the ZrTe$_2$ shows 30$^\circ$ twin domains, consistent with a trigonal material grown on the cubic FeTe~(See SM~\cite{Islam_Supp}).
    
The electronic band structure obtained from \textit{in situ} angle resolved photo-emission spectroscopy (ARPES) for a  4UC ZrTe$_2$/FeTe hybrid is shown in Fig.~\ref{1}(h) (See SM~\cite{Islam_Supp}). The corresponding second derivative spectra is shown in Fig.~\ref{1}(i). The ARPES data clearly demonstrates linearly dispersing DSM bands, as observed in prior studies of \ZrTe~grown on other substrates and buffer layers~\cite{tsipas2018massless,muhammad2019transition,ou2022zrte2}, with the chemical potential located below the Dirac point, consistent with transport measurements (See SM~\cite{Islam_Supp}).
     
We now discuss electrical transport measurements, conducted on mechanically scratched Hall bars with length=$1$~mm and width=$0.5$~mm. Fig.~\ref{2}(a) shows the temperature variation of the sample resistance for 2 UC, 6 UC, 8 UC \ZrTe~ and 1 UC CrTe$_2$/6 UC ~\ZrTe~ grown on 35 UC FeTe. All these samples show a transition from a normal metal to a superconducting state with a minimum resistance smaller than our measurement limits ($R \lesssim 1$~m$\Omega$), except for the 2 UC sample whose resistance does not fully vanish at $T=2$~K. We attribute the latter to inhomogeneity and possibly incomplete coverage of the FeTe with ~\ZrTe~over the scale of the Hall bar device. We observe a distinctive resistance peak around $T = 60$~K in all samples. This is more pronounced in the 2 UC sample and broadens as we increase \ZrTe~ thickness. This feature is consistent with critical spin scattering at the N{\'e}el temperature of FeTe as it transitions from a paramagnet to an AFM~\cite{liu2009charge,lee2019spin}. In light of recent insights about the tension between antiferromagnetism and superconductivity \cite{yan2026stoichiometric}, we hypothesize that our as-grown FeTe films likely have Fe intersitials, resulting in AFM order. The transition to a superconducting phase then likely occurs only near the interface where Fe interstitials are removed during the overgrowth of \ZrTe. Proving this hypothesis will require additional measurements such as more careful HRTEM or cross-sectional scanning tunneling microscopy. We also observe an excess resistance anomaly before the superconducting transition; we attribute this to non-equilibrium effects at a normal-superconductor interface in the presence of inhomogeneities~\cite{park1995resistance,zhang2013metal,si2010superconductivity}. The sample-to-sample variability in the position of the resistance peak, indicates that its energy scale is not uniquely set by the CrTe$_2$-induced exchange field, but likely also reflects strain- and stoichiometry-related shifts of the FeTe N{\'e}el temperature, with a possible additional contribution from parallel conduction through the CrTe$_2$ layer. We also note that the CrTe$_2$ induces a spin splitting on the order of meV throughout the ZrTe$_2$ layer which is comparable to that of the superconducting gap at the ZrTe$_2$/FeTe interface (See SM~\cite{Islam_Supp}).
    
To gain deeper insights into the nature of the superconducting state, we measured the temperature dependent resistance in different magnetic fields, applied perpendicular and parallel to the sample plane; this is shown in Figs.~\ref{2}(b) and ~\ref{2}(c), respectively, for the 6 UC \ZrTe~ sample (See SM~\cite{Islam_Supp}). The general characteristics observed here are superficially similar to the behavior reported in prior studies of emergent superconductivity in FeTe-based heterostructures~\cite{he2014two,Yasuda_NComm,liang2020studies,yao2021hybrid,yao2022superconducting,tkavc2023multiphase,yao2024mystery,yi2023dirac,hemianCBST,yuan2024coexistence,Yan_2025}. The upper critical field is clearly very large ($\geq 14$~T) for both the in-plane and out-of-plane applied field directions. The superconducting transition exhibits a pronounced anisotropy with respect to the direction of the applied external magnetic field: the broadening of the transition is significantly weaker when the magnetic field is applied parallel to the interface, compared to perpendicular direction. We also measure the non-linear current-voltage characteristic (IVC) of the films close to the transition and find a hysteretic behavior with $T$-dependent switching ($I_c$) and retrapping ($I_r$) currents,  with $|I_r|< |I_c|$ (Fig.~\ref{2}(d)), attributed to the formation of hot spots (See SM~\cite{Islam_Supp} for $T$-dependence of $I_c$). We next extract the upper critical field ($H_{c2}$) as a function of reduced temperature ($T/T_c$) for both perpendicular and parallel magnetic field directions (Fig.~\ref{3}(a)). We define the transition temperature ($T_c$) as the temperature at which $R_{xx}$ drops to 50\% of the value recorded at $T=30$~K. We use the weak coupling BCS theory for 2D superconductors ($T_c/T_f<0.005$, $T_f$ being the Fermi temperature~\cite{cao2018unconventional,park2021tunable}) to fit the critical field vs. reduced temperature ($T/T_c)$ and extract the zero-temperature coherence length ($\xi_{0}$) and superconducting length ($d_{sc}$)~\cite{he2014two}. Here, the perpendicular critical-field ($H_{c2,\perp}$) dependence is given by
    
\begin{equation}
\mu_0 {H}_{c2,\perp}\left(T\right)=\frac{{\Phi }_{0}}{2\pi \xi_{0}^{2}}{\left(1-T/{T}_{C}\right)}
\end{equation} whereas the parallel critical field ($H_{c2,\parallel}$) dependence is given by
\begin{equation}
\mu_0{H}_{c2,\parallel}\left(T\right)=\frac{\sqrt{3}{\Phi }_{0}}{\pi {\xi_0d_{sc}}}{\left(1-T/{T}_{C}\right)^{1/2}}
\end{equation}
Here, $\Phi_0$ is the magnetic flux quantum. From reasonable fits to our data (Fig.~\ref{3}(a)), we estimate $\xi_0=1.9$~nm and superconducting length $d_{sc}=12.9$~nm. 
In all samples studied here, the experimentally extracted in-plane critical field extrapolated to $T=0$~K is twice the value expected theoretically for Pauli pair breaking~\cite{chandrasekhar1962note,clogston1962upper}. The apparently ubiquitous observation of this violation in FeTe-based heterostructures raises interesting questions regarding the pairing symmetry but is beyond the scope of the present work.

Since superconductivity in our heterostructures is assumed to be 2D in nature, arising at the interface with FeTe, as hypothesized in other FeTe-based heterostructures \cite{he2014two,liang2020studies,hemianCBST}, we analyze the temperature dependence of sample resistance according to the Berezinskii–Kosterlitz–Thouless (BKT) model for 2D superconductors. This is best described using the Halperin-Nelson equation, where $R(T)\propto R_0 \exp^{(-b/(T-T_{\rm{BKT}})^\frac{1}{2})}$. Here, $R_0$ and $b$ are material-specific constants~\cite{kosterlitz1973rdering,kosterlitz1974critical,he2014two,halperin1979resistive,lin2012suppression}. Figure~\ref{3}(b) shows that this model works reasonably well for the 6 UC samples, generating $T_{BKT}=9.6$~K. Within the BKT framework, the IV relationship follows a power-law dependence ($V\propto I^\alpha$), where $\alpha$ depends on the temperature~\cite{venditti2019nonlinear,kumar2023flux}. $\alpha=3$ is a hallmark of the BKT transition arising due to the onset of free vortex-antivortex motion. We fit the IVC within the BKT framework (Fig.~\ref{3}(c)) and show the corresponding magnitude of $\alpha$ in Fig.~\ref{3}(d). Above the transition, the IVC is linear while the value of $\alpha$ increases rapidly below $T_c$ to $13$ at $T=2$~K. At $T=9.5$~K, $\alpha=3$, which corresponds to $T_{\rm{BKT}}$. This is in excellent agreement with the analysis of the $R$ vs $T$ data, providing strong support for the 2D nature of superconductivity in these samples. 
    
We next examine non-reciprocal transport to explore the interaction between superconductivity and topological states. When an in-plane magnetic field is applied (Fig.~\ref{4}(a)), the longitudinal voltage $V_{xx}$ under broken inversion symmetry is phenomenologically given by 
\begin{equation}
V_{xx} = R^\omega I(1 + \gamma\mu_0 \,({H} \times \hat z) \cdot {I})    
\end{equation}
Here, $R^\omega$ is the first-harmonic longitudinal resistance
and $\gamma$ is the magneto-chiral anisotropy coefficient which quantifies the strength of the non-reciprocal charge transport. When the current is perpendicular to the applied in-plane field, Eqn. 3 reduces to $V_{xx}=R^\omega I+ \gamma\mu_0 R^{\omega} HI^2$.
We use the second term to extract $\gamma$ by measuring the second harmonic signal under applied ac-current due to its quadratic nature, where $R^{2\omega } = \frac{{R^\omega}}{{\sqrt 2 }}\gamma\mu_0 HI$. Experimentally, $R^{2\omega}$ is measured as a function of magnetic field at different temperatures (See SM~\cite{Islam_Supp}). $\gamma$ can be extracted from the slope of ${R^{2\omega}}/{R^\omega}$ vs. $H$. We extract $\gamma$ as a function of temperature (Fig.~\ref{4}(b)) and divide the response into three regions: normal, intermediate, and superconducting. In the superconducting region, both $R^{2\omega}$ and $R^\omega$ vanish, whereas in the normal region, $R^{2\omega}$ becomes negligible. In the intermediate region where $T_{BKT}<T<T_{c,onset}$, the extracted value of $\gamma$ appears to diverge with a functional form $\propto (T-T_{BKT})^{-3/2}$. In the 6 UC sample, the magnitude of $\gamma$, normalized with current density, is comparable to that observed in Bi$_2$Te$_3$/FeTe hybrids with a maximum of $6$~A$^{-1}$T$^{-1}$m. The enhancement of the magneto-chiral anisotropy arises due to shift of the energy scale from the Fermi energy ($E_f$) to the superconducting gap ($\Delta_{SC}$) which is smaller~\cite{hoshino2018nonreciprocal,rikken2005magnetoelectric}.  
We also observe a threefold enhancement in the magnitude of $\gamma$ when a monolayer of CrTe$_2$, a 2D vdW ferromagnet with perpendicular magnetic anisotropy, is grown on top of the STO/FeTe/ZrTe$_2$(6 UC) heterostructure (Fig.~\ref{4}(b) (See SM~\cite{Islam_Supp}). We are tempted to attribute this to the removal of time-reversal symmetry, as in prior studies of ferromagnets directly interfaced with FeTe \cite{Yan_2017}, but note that the ferromagnetic layer in our samples is $\sim 4$~nm from the FeTe interface where superconductivity presumably resides. Another aspect of non-reciprocity is the superconducting diode effect, which manifests as an asymmetry in the critical current in the positive and negative current branches~\cite{ando2020observation,hou2023ubiquitous} (Fig.~\ref{4}(c) and SM {\cite{Islam_Supp}). In Fig.~\ref{4}(d), we plot the difference between the magnitude of the critical current between positive and negative sweep directions, $\Delta I_c= I_{c}^{+}-|I_{c}^{-}|$ as a function of magnetic field for the STO/FeTe/ZrTe$_2$(6 UC)/CrTe$_2$(1 UC) heterostructure. Here, the midpoint of the $V–I$ curve is defined as the critical current with $I_{c}^{+}$ and $|I_{c}^{-}|$ denoting the corresponding values for positive and negative current sweeps~\cite{ando2020observation}. While $\Delta I_c$ decreases as $\mu_0 H_{c2,\parallel}$ increases, the efficiency of the diode effect, $\eta = \frac{I^+_c-|I^-_c|}{I^+_c+|I^-_c|}$ shows a moderate increase with $\mu_0 H$ and can be as high as $29$\%, which is similar to the highest reported values in the literature~\cite{ma2025superconducting}. We note here that the Pauli limit violation combined with the high diode efficiency are suggestive of unconventional or finite-momentum pairing in the interfacial superconducting state, but the current measurements do not permit a distinction between spin-orbit-coupled superconductor with broken inversion and time-reversal symmetry, and a genuine finite-momentum pairing state~\cite{nadeem2023superconducting}. We note that we could not reliably extract the critical current for magnetic fields below $3$~T over the temperature range investigated; also, corresponding measurements under negative magnetic fields are not available.

In the framework discussed by Yuan and Fu~\cite{yuan2022supercurrent}, a superconducting diode effect arises when inversion symmetry breaking and time-reversal symmetry breaking allow odd-in-momentum terms in the Ginzburg-Landau free energy of a current-carrying superconducting condensate. A supercurrent corresponds to a finite condensate momentum $q$, and in a spin-orbit-coupled noncentrosymmetric superconductor the free energy may contain terms of the form 
\begin{equation}
    F(\vec{q}, \Delta) = \bigl\{ t + a_{0} q^{2} - b_{0} \, \vec{q} \cdot (\vec{h}_{\mathrm{eff}} \times \hat{z}) + b_{1} q^{2} \vec{q} \cdot (\vec{h}_{\mathrm{eff}} \times \hat{z}) \bigr\} |\Delta|^{2} + \frac{\beta}{2} |\Delta|^{4}
    \label{eq:placeholder_label}
\end{equation}

Here $\hat{z}$ is the polar axis of the ZrTe$_2$/FeTe interface, ${\vec{h}}_{\mathrm{eff}}$ is the effective time-reversal-breaking field, and $\Delta$ is the superconducting order parameter. The term $q^2\vec{q}\cdot\left({\vec{h}}_{\mathrm{eff}}\times\hat{z}\right)$ is odd in q and skews the depairing free energy for positive and negative current directions. This produces unequal critical currents in opposite directions, which is the superconducting diode effect. Thus, the diode response is controlled not only by the existence of inversion breaking, but also by the magnitude of the effective time-reversal-breaking field acting on the spin-orbit-coupled superconducting condensate. In our heterostructure, the effective field can be written phenomenologically as:
${\vec{h}}_{\mathrm{eff}}={\vec{h}}_{\mathrm{ex}}+{\vec{h}}_{\mathrm{prox}}$
The first term ${\vec{h}}_{\mathrm{ex}}$ is the ordinary Zeeman coupling to the applied magnetic field. The second term ${\vec{h}}_{\mathrm{prox}}$ represents the exchange-proximity contribution from the CrTe$_2$ layer. Thus, the CrTe$_2$ cap does not provide the inversion symmetry breaking; that is already present at the ZrTe$_2$/FeTe interface. Instead, CrTe$_2$ provides an additional exchange field that adds to the applied-field Zeeman term. Since the odd-in-$q$ terms responsible for asymmetric depairing scale with ${\vec{h}}_{\mathrm{eff}}$, this additional exchange contribution provides a natural microscopic mechanism for the enhanced superconducting diode response observed after CrTe$_2$ capping. To further support this interpretation, we have added first-principles calculations of the proximity effect in the SI (Fig. S13-S15). Specifically, we calculate the spin density in the CrTe$_2$/ZrTe$_2$ heterostructure and find that the magnetic polarization associated with CrTe$_2$ induces a finite spin density in the adjacent ZrTe$_2$ layer. This result provides microscopic support for the phenomenological exchange-proximity term ${\vec{h}}_{\mathrm{prox}}$ used above. 
We further note that although the calculated spin polarization decays rapidly away from the CrTe$_2$/ZrTe$_2$ interface, this small magnetic proximity gives rise to a meV-scale spin splitting throughout the ZrTe$_2$ layer (See SM~\cite{Islam_Supp}). This splitting is not negligible compared to the energy scale of the superconducting gap at the ZrTe$_2$/FeTe interface.

The detailed underlying physical origin of interface-induced superconductivity in FeTe-based hybrids remains unresolved~\cite{yao2024mystery}, although a recent report of stoichiometric FeTe being a superconductor suggests that capping as-grown FeTe films with other tellurides can effectively anneal away interstitial defects ~\cite{yan2026stoichiometric}. The interplay between antiferromagnetism and superconductivity has been investigated in ~\cite{balakrishnan2025complex}, where it has been observed that antiferromagnetism in FeTe is suppressed in (BiSb)$_2$Te$_3$/FeTe and Cr-(BiSb)$_2$Te$_3$/FeTe hybrids, yet co-exists with superconductivity when interfaced with MnBi$_2$Te$_4$. To gain further insight into the superconductivity observed in our experiments, we perform first-principles calculations based on density functional theory (DFT) (See Supplemental Material for details \cite{Islam_Supp}). As a first step, we investigate the role of the ZrTe$_2$ layer by calculating the interfacial charge transfer in ZrTe$_2$/FeTe heterostructures (Fig.~S12c in SM). The results reveal an accumulation of electrons in the ZrTe$_2$ layer, indicating hole doping in FeTe near the interface. This finding is further supported by additional DFT calculations for isolated bilayer ZrTe$_2$ and FeTe, which show that 2 UC ZrTe$_2$ has a higher work function ($4.83$~eV) than 2 UC FeTe ($4.25$~eV). Next, we examine the relative stability of the bicollinear and stripe antiferromagnetic phases in 2 UC FeTe as a function of hole doping (Fig.~S12d in SM). In the pristine case, although the bicollinear AFM is the experimental ground state of FeTe, the total energy of the stripe AFM phase is calculated to be lower than that of the bicollinear AFM, consistent with previous theoretical reports~\cite{koteski2017first}. We note that the relative stability of magnetic phases can be sensitive to various computational details. Therefore, in the following discussion, we focus on the trend with hole doping based on our theoretical results. Interestingly, we find that hole doping stabilizes the bicollinear AFM, suggesting that this phase remains the ground state in the experimentally studied ZrTe$_2$/FeTe heterostructure. Therefore, the interfacial superconductivity observed in the ZrTe$_2$/FeTe heterostructure may coexist with the bicollinear antiferromagnetic order, similar to previous observations in FeTe/Bi$_2$Te$_3$~\cite{he2014two,yi2023dirac} and FeTe/MnBi$_2$Te$_4$~\cite{yuan2024coexistence} heterostructures. These results indicate that the primary role of ZrTe$_2$ is to induce charge transfer and modulate the magnetic order of FeTe, providing new insight into the origin of the mysterious interfacial superconductivity in FeTe-based heterostructures.

In conclusion, we have demonstrated an epitaxial vdW platform to induce superconductivity in a DSM, ZrTe$_2$, while simultaneously interfacing it with an AFM (FeTe) and a 2D vdW FM (CrTe$_2$). Through detailed electrical transport measurements, we demonstrate that 2D superconductivity emerges at the interface below a critical temperature of $12$~K. This is an unusual superconducting state that supports non-reciprocal transport, as revealed through magneto-chiral anisotropy and the superconducting diode effect. The especially high efficiency ($29$\%) of the latter in hybrid CrTe$_2$/ZrTe$_2$/FeTe heterostructures promises an attractive wafer scale vdW platform for superconducting electronics. Although the current transport data reflect the response of the heterostructure, the state of $ZrTe_2$ Dirac fermions in the superconducting state remains an open question that needs to be addressed. Beyond the present study, these hybrid epitaxial heterostructures provide a potentially versatile platform for exploring topological superconductivity and its interaction with magnetism.

    \section{Acknowledgments}

    This work was supported primarily by the Penn State Two-Dimensional Crystal Consortium Materials Innovation Platform (2DCC-MIP) under NSF Grant No. DMR-2039351 using the multi-module UHV MBE facility (DOI: 10.60551/jb0f-5784) and by NSF Grant No. DMR-2407130. We also acknowledge the low-temperature transport facilities (DOI: 10.60551/rxfx-9h58) provided by the Penn State Materials Research Science and Engineering Center under award NSF-DMR 2011839. K.D.H. and D.R.H. acknowledge support through startup funds from the Penn State Eberly College of Science, Department of Chemistry, College of Earth and Mineral Sciences, Department of Materials Science and Engineering, and Materials Research Institute. The authors also acknowledge the use of the Penn State Materials Characterization Lab. Finally, the authors thank H. Yi, C.-Z. Chang, and C.-X. Liu for helpful discussions.

\bibliography{ref.bib}

@article{he2014two,
  title={Two-dimensional superconductivity at the interface of a {Bi$_2$Te$_3$/FeTe} heterostructure},
  author={He, Qing Lin and Liu, Hongchao and He, Mingquan and Lai, Ying Hoi and He, Hongtao and Wang, Gan and Law, Kam Tuen and Lortz, Rolf and Wang, Jiannong and Sou, Iam Keong},
  journal={Nat. Commun.},
  volume={5},
  number={1},
  pages={4247},
  year={2014},
  publisher={Nature Publishing Group UK London},
  doi={https://www.nature.com/articles/ncomms5247}
}

@article{yuan2024coexistence,
  title={Coexistence of Superconductivity and Antiferromagnetism in Topological Magnet MnBi$_2$Te$_4$ Films},
  author={Yuan, Wei and Yan, Zi-Jie and Yi, Hemian and Wang, Zihao and Paolini, Stephen and Zhao, Yi-Fan and Zhou, Lingjie and Wang, Annie G and Wang, Ke and Prokscha, Thomas and Zaher Salman and 
Andreas SuterPurnima and P. Balakrishnan and Alexander J. Grutter and Laurel E. Winter and John Singleton and Moses H. W. Chan and Cui-Zu Chang},
  journal={Nano Lett.},
volume={24},
pages={7962},
  year={2024},
  publisher={ACS Publications},
  doi={https://pubs.acs.org/doi/10.1021/acs.nanolett.4c01407}
}

@article{hemianCBST,
author = {Hemian Yi  and Yi-Fan Zhao  and Ying-Ting Chan  and Jiaqi Cai  and Ruobing Mei  and Xianxin Wu  and Zi-Jie Yan  and Ling-Jie Zhou  and Ruoxi Zhang  and Zihao Wang  and Stephen Paolini  and Run Xiao  and Ke Wang  and Anthony R. Richardella  and John Singleton  and Laurel E. Winter  and Thomas Prokscha  and Zaher Salman  and Andreas Suter  and Purnima P. Balakrishnan  and Alexander J. Grutter  and Moses H. W. Chan  and Nitin Samarth  and Xiaodong Xu  and Weida Wu  and Chao-Xing Liu  and Cui-Zu Chang },
title = {Interface-induced superconductivity in magnetic topological insulators},
journal = {Science},
volume = {383},
number = {6683},
pages = {634-639},
year = {2024},
doi = {10.1126/science.adk1270},
doi = {https://www.science.org/doi/abs/10.1126/science.adk1270}
}

@article{tkavc2023multiphase,
  title={Multiphase superconductivity at the interface between ultrathin FeTe islands and Bi$_2$Te$_3$},
  author={Tk{\'a}{\v{c}}, Vladimir and Vorobiov, Serhii and Baloh, Pavlo and Vondracek, Martin and Springholz, Gunther and Carva, Karel and Szab{\'o}, Pavol and Hofmann, Philip and Honolka, Jan},
  journal = {npj 2D Mater. Appl.},
  volume = {8},
  number = {52},
  year={2023},
  doi={https://www.nature.com/articles/s41699-024-00480-x}
}

@article{yao2024mystery,
  title={Mystery of superconductivity in FeTe films and the role of neighboring layers},
  author={Yao, Xiong and Yi, Hee Taek and Jain, Deepti and Yuan, Xiaoyu and Oh, Seongshik},
  journal={arXiv:2410.17671},
  year={2024},
  doi={https://arxiv.org/abs/2410.17671}
}

@article{yao2021hybrid,
  title={Hybrid symmetry epitaxy of the superconducting Fe(Te,Se) film on a topological insulator},
  author={Yao, Xiong and Brahlek, Matthew and Yi, Hee Taek and Jain, Deepti and Mazza, Alessandro R and Han, Myung-Geun and Oh, Seongshik},
  journal={Nano Lett.},
  volume={21},
  number={15},
  pages={6518--6524},
  year={2021},
  publisher={ACS Publications},
  doi={https://pubs.acs.org/doi/full/10.1021/acs.nanolett.1c01703}
}

@article{yao2022superconducting,
  title={Superconducting fourfold Fe (Te, Se) film on sixfold magnetic MnTe via hybrid symmetry epitaxy},
  author={Yao, Xiong and Mazza, Alessandro R and Han, Myung-Geun and Yi, Hee Taek and Jain, Deepti and Brahlek, Matthew and Oh, Seongshik},
  journal={Nano Lett.},
  volume={22},
  number={18},
  pages={7522--7526},
  year={2022},
  publisher={ACS Publications},
doi={https://pubs.acs.org/doi/full/10.1021/acs.nanolett.2c02510}
}

@article{yi2023dirac,
  title={Dirac-fermion-assisted interfacial superconductivity in epitaxial topological-insulator/iron-chalcogenide heterostructures},
  author={Yi, Hemian and Hu, Lun-Hui and Zhao, Yi-Fan and Zhou, Ling-Jie and Yan, Zi-Jie and Zhang, Ruoxi and Yuan, Wei and Wang, Zihao and Wang, Ke and Hickey, Danielle Reifsnyder and others},
  journal={Nat. Commun.},
  volume={14},
  number={1},
  pages={7119},
  year={2023},
  publisher={Nature Publishing Group UK London},
  doi={https://www.nature.com/articles/s41467-023-42902-2}
}

@article{ou2022zrte2,
  title={ZrTe$_2$/CrTe$_2$: an epitaxial van der Waals platform for spintronics},
  author={Ou, Yongxi and Yanez, Wilson and Xiao, Run and Stanley, Max and Ghosh, Supriya and Zheng, Boyang and Jiang, Wei and Huang, Yu-Sheng and Pillsbury, Timothy and Richardella, Anthony and others},
  journal={Nat. Commun.},
  volume={13},
  number={1},
  pages={2972},
  year={2022},
  publisher={Nature Publishing Group UK London},
  doi={https://www.nature.com/articles/s41467-022-30738-1}
}

@article{liang2020studies,
  title={Studies on the origin of the interfacial superconductivity of {Sb$_2$Te$_3$/Fe$_{1+y}$Te} heterostructures},
  author={Liang, Jing and Zhang, Yu Jun and Yao, Xiong and Li, Hui and Li, Zi-Xiang and Wang, Jiannong and Chen, Yuanzhen and Sou, Iam Keong},
  journal={Proc. Natl. Acad. Sci.},
  volume={117},
  number={1},
  pages={221--227},
  year={2020},
  publisher={National Acad Sciences},
  doi={https://www.pnas.org/doi/full/10.1073/pnas.1914534117}
}

@article{huang2019proximity,
  title={Proximity-induced surface superconductivity in Dirac semimetal Cd$_3$As$_2$},
  author={Huang, Ce and Zhou, Benjamin T and Zhang, Huiqin and Yang, Bingjia and Liu, Ran and Wang, Hanwen and Wan, Yimin and Huang, Ke and Liao, Zhiming and Zhang, Enze and  Liu, S and Deng, Q and Chen, Y and Han, X and Zou, J and Lin, X and Han, Z and Wang, Y and Law, KT and Xiu, F},
  journal={Nat. Commun.},
  volume={10},
  number={1},
  pages={2217},
  year={2019},
  publisher={Nature Publishing Group UK London},
  doi={https://www.nature.com/articles/s41467-019-10233-w}
}

@article{rashidi2023induced,
  title={Induced superconductivity in the two-dimensional topological insulator phase of cadmium arsenide},
  author={Rashidi, Arman and Kealhofer, Robert and Lygo, Alexander C and Huang, Victor and Stemmer, Susanne},
  journal={APL Mater.},
  volume={11},
  number={4},
  year={2023},
  publisher={AIP Publishing},
  doi={https://pubs.aip.org/aip/apm/article/11/4/041117/2878523/Induced-superconductivity-in-the-two-dimensional}
}

@article{kobayashi2015topological,
  title={Topological superconductivity in Dirac semimetals},
  author={Kobayashi, Shingo and Sato, Masatoshi},
  journal={Phys. Rev. Lett.},
  volume={115},
  number={18},
  pages={187001},
  year={2015},
  publisher={APS},
  doi={https://journals.aps.org/prl/abstract/10.1103/PhysRevLett.115.187001}
}

@article{bobrow2020monopole,
  title={Monopole charge density wave states in Weyl semimetals},
  author={Bobrow, Eric and Sun, Canon and Li, Yi},
  journal={Phys. Rev. Research},
  volume={2},
  number={1},
  pages={012078},
  year={2020},
  publisher={APS},
  doi={https://journals.aps.org/prresearch/abstract/10.1103/PhysRevResearch.2.012078}
}

@article{li2018topological,
  title={Topological nodal Cooper pairing in doped Weyl metals},
  author={Li, Yi and Haldane, FDM},
  journal={Phys. Rev. Lett.},
  volume={120},
  number={6},
  pages={067003},
  year={2018},
  publisher={APS},
  doi={https://journals.aps.org/prl/abstract/10.1103/PhysRevLett.120.067003}
}

@article{sun2019vortices,
  title={Vortices in a monopole superconducting Weyl semi-metal},
  author={Sun, Canon and Lee, Shu-Ping and Li, Yi},
  journal={arXiv:1909.04179},
  year={2019},
  doi={https://arxiv.org/abs/1909.04179}
}

@article{armitage2018weyl,
  title={Weyl and Dirac semimetals in three-dimensional solids},
  author={Armitage, NP and Mele, EJ and Vishwanath, Ashvin},
  journal={Rev. Mod. Phys.},
  volume={90},
  number={1},
  pages={015001},
  year={2018},
  publisher={APS},
doi={https://journals.aps.org/rmp/abstract/10.1103/RevModPhys.90.015001}
}

@article{Yan_2017,
author = {Yan, Binghai. and Felser, Claudia.},
year = {2016},
Title  = {Topological semimetals: {W}eyl semimetals},
Journal  = {Ann. Rev. Cond. Matt. Phys.},
SP  = {22377},
Pages  = {337-354},
doi  = {https://doi.org/10.1146/annurev-conmatphys-031016-025458}
}

@article{borisenko2014experimental,
  title={Experimental realization of a three-dimensional {Dirac} semimetal},
  author={Borisenko, Sergey and Gibson, Quinn and Evtushinsky, Danil and Zabolotnyy, Volodymyr and B{\"u}chner, Bernd and Cava, Robert J},
  journal={Phys. Rev. Lett.},
  volume={113},
  number={2},
  pages={027603},
  year={2014},
  publisher={APS},
  doi={https://journals.aps.org/prl/abstract/10.1103/PhysRevLett.113.027603}
}

@article{burkov2016topological,
  title={Topological semimetals},
  author={Burkov, A. A.},
  journal={Nat. Mater.},
  volume={15},
  number={11},
  pages={1145--1148},
  year={2016},
  publisher={Nature Publishing Group},
  doi={https://nature.com/articles/nmat4788}
}

@article{yan2017topological,
  title={Topological materials: {Weyl} semimetals},
  author={Yan, Binghai and Felser, Claudia},
  journal={Annu. Rev. Condens. Matter Phys.},
  volume={8},
  pages={337--354},
  year={2017},
  publisher={Annual Reviews},
  doi={https://www.annualreviews.org/content/journals/10.1146/annurev-conmatphys-031016-025458}
}

@article{bednik2015superconductivity,
  title={Superconductivity in Weyl metals},
  author={Bednik, Grigory and Zyuzin, AA and Burkov, AA},
  journal={Phys. Rev. B},
  volume={92},
  number={3},
  pages={035153},
  year={2015},
  publisher={APS},
  doi={https://journals.aps.org/prb/abstract/10.1103/PhysRevB.92.035153}
}

@article{subedi2008density,
  title={Density functional study of FeS, FeSe, and FeTe: Electronic structure, magnetism, phonons, and superconductivity},
  author={Subedi, Alaska and Zhang, Lijun and Singh, David J and Du, Mao-Hua},
  journal={Phys. Rev. B},
  volume={78},
  number={13},
  pages={134514},
  year={2008},
  publisher={APS},
  doi={https://journals.aps.org/prb/abstract/10.1103/PhysRevB.78.134514}
}

@article{chen2009electronic,
  title={Electronic properties of single-crystalline Fe$_{1.05}$Te and Fe$_{1.03}$Se$_{0.30}$Te$_{0.70}$},
  author={Chen, GF and Chen, ZG and Dong, J and Hu, WZ and Li, G and Zhang, XD and Zheng, P and Luo, JL and Wang, NL},
  journal={Phys. Rev. B},
  volume={79},
  number={14},
  pages={140509},
  year={2009},
  publisher={APS},
  doi={https://journals.aps.org/prb/abstract/10.1103/PhysRevB.79.140509}
}

@article{hou2023ubiquitous,
  title={Ubiquitous superconducting diode effect in superconductor thin films},
  author={Hou, Yasen and Nichele, Fabrizio and Chi, Hang and Lodesani, Alessandro and Wu, Yingying and Ritter, Markus F and Haxell, Daniel Z and Davydova, Margarita and Ili{\'c}, Stefan and Glezakou-Elbert, Ourania and  Varambally, Amith and Bergeret, F. Sebastian and Kamra, Akashdeep and Fu, Liang and Lee, Patrick A. and Moodera, Jagadeesh S.},
  journal={Phys. Rev. Lett.},
  volume={131},
  number={2},
  pages={027001},
  year={2023},
  publisher={APS},
  doi={https://journals.aps.org/prl/abstract/10.1103/PhysRevLett.131.027001}
}

@article{ando2020observation,
  title={Observation of superconducting diode effect},
  author={Ando, Fuyuki and Miyasaka, Yuta and Li, Tian and Ishizuka, Jun and Arakawa, Tomonori and Shiota, Yoichi and Moriyama, Takahiro and Yanase, Youichi and Ono, Teruo},
  journal={Nature},
  volume={584},
  number={7821},
  pages={373--376},
  year={2020},
  publisher={Nature Publishing Group UK London},
  doi={https://www.nature.com/articles/s41586-020-2590-4}
}

@article{kosterlitz1973rdering,
  title={Ordering, metastability and phase transitions in two-dimensional systems},
  author={Kosterlitz, John Michael and Thouless, David James},
  journal={J. Phys. C: Solid State Phys},
  volume={6},
  pages={1181},
  year={1973},
  publisher={IOP Publishing},
  doi={https://iopscience.iop.org/article/10.1088/0022-3719/6/7/010}
}

@article{kosterlitz1974critical,
  title={The critical properties of the two-dimensional xy model},
  author={Kosterlitz, J Michael},
  journal={J. Phys. C: Solid State Phys},
  volume={7},
  number={6},
  pages={1046},
  year={1974},
  publisher={IOP Publishing},
  doi={https://iopscience.iop.org/article/10.1088/0022-3719/7/6/005}
}

@article{fu2008superconducting,
  title={Superconducting Proximity Effect and Majorana Fermions at the Surface of a Topological Insulator},
  author={Fu, Liang and Kane, Charles L},
  journal={Phys. Rev. Lett.},
  volume={100},
  number={9},
  pages={096407},
  year={2008},
  publisher={APS},
  doi={https://journals.aps.org/prl/abstract/10.1103/PhysRevLett.100.096407}
}

@article{qi2011topological,
  title={Topological insulators and superconductors},
  author={Qi, Xiao-Liang and Zhang, Shou-Cheng},
  journal={Rev. Mod. Phys.},
  volume={83},
  number={4},
  pages={1057--1110},
  year={2011},
  publisher={APS},
  doi={https://journals.aps.org/rmp/abstract/10.1103/RevModPhys.83.1057}
}

@article{ong2021experimental,
  title={Experimental signatures of the chiral anomaly in Dirac--Weyl semimetals},
  author={Ong, NP and Liang, Sihang},
  journal={Nat. Rev. Phys.},
  volume={3},
  number={6},
  pages={394--404},
  year={2021},
  publisher={Nature Publishing Group UK London},
  doi={https://www.nature.com/articles/s42254-021-00310-9}
}

@article{xiong2015evidence,
  title={Evidence for the chiral anomaly in the Dirac semimetal Na$_3$Bi},
  author={Xiong, Jun and Kushwaha, Satya K and Liang, Tian and Krizan, Jason W and Hirschberger, Max and Wang, Wudi and Cava, Robert Joseph and Ong, Nai Phuan},
  journal={Science},
  volume={350},
  number={6259},
  pages={413--416},
  year={2015},
  publisher={American Association for the Advancement of Science},
  doi={https://www.science.org/doi/10.1126/science.aac6089}
}

@article{shekhar2015extremely,
  title={Extremely large magnetoresistance and ultrahigh mobility in the topological Weyl semimetal candidate NbP},
  author={Shekhar, Chandra and Nayak, Ajaya K and Sun, Yan and Schmidt, Marcus and Nicklas, Michael and Leermakers, Inge and Zeitler, Uli and Skourski, Yurii and Wosnitza, Jochen and Liu, Zhongkai and others},
  journal={Nat. Phys.},
  volume={11},
  number={8},
  pages={645--649},
  year={2015},
  publisher={Nature Publishing Group UK London}
}

@article{halperin1979resistive,
  title={Resistive transition in superconducting films},
  author={Halperin, BI and Nelson, David R},
  journal={J. Low Temp. Phys.},
  volume={36},
  pages={599--616},
  year={1979},
  publisher={Springer},
  doi={https://link.springer.com/article/10.1007/BF00116988}
}

@article{lin2012suppression,
  title={Suppression of the Berezinskii-Kosterlitz-Thouless Transition in 2D Superconductors by Macroscopic Quantum Tunneling},
  author={Lin, Yen-Hsiang and Nelson, J and Goldman, AM},
  journal={Phys. Rev. Lett.},
  volume={109},
  number={1},
  pages={017002},
  year={2012},
  publisher={APS},
  doi={https://journals.aps.org/prl/abstract/10.1103/PhysRevLett.109.017002}
}

@article{chandrasekhar1962note,
  title={A note on the maximum critical field of high-field superconductors},
  author={Chandrasekhar, BS},
  journal={Appl. Phys. Letters},
  volume={1},
  year={1962},
  publisher={Westinghouse Research Labs., Pittsburgh},
  doi={https://pubs.aip.org/aip/apl/article/1/1/7/976920/A-NOTE-ON-THE-MAXIMUM-CRITICAL-FIELD-OF-HIGH-FIELD}
}

@article{clogston1962upper,
  title={Upper limit for the critical field in hard superconductors},
  author={Clogston, Albert M},
  journal={Phys. Rev. Lett.},
  volume={9},
  number={6},
  pages={266},
  year={1962},
  publisher={APS},
  doi={https://journals.aps.org/prl/abstract/10.1103/PhysRevLett.9.266}
}

@article{hoshino2018nonreciprocal,
  title={Nonreciprocal charge transport in two-dimensional noncentrosymmetric superconductors},
  author={Hoshino, Shintaro and Wakatsuki, Ryohei and Hamamoto, Keita and Nagaosa, Naoto},
  journal={Phys. Rev. B},
  volume={98},
  number={5},
  pages={054510},
  year={2018},
  publisher={APS},
  doi={https://journals.aps.org/prb/pdf/10.1103/PhysRevB.98.054510}
}

@article{rikken2005magnetoelectric,
  title={Magnetoelectric anisotropy in diffusive transport},
  author={Rikken, GLJA and Wyder, P},
  journal={Phys. Rev. Lett.},
  volume={94},
  number={1},
  pages={016601},
  year={2005},
  publisher={APS},
 doi={https://journals.aps.org/prl/abstract/10.1103/PhysRevLett.94.016601}
}

@article{rashidi2024vortex,
  title={Vortex-induced anomalies in the superconducting quantum interference patterns of topological insulator Josephson junctions},
  author={Rashidi, Arman and Huynh, William and Guo, Binghao and Ahadi, Sina and Stemmer, Susanne},
  journal={npj Quantum Mater.},
  volume={9},
  number={1},
  pages={70},
  year={2024},
  publisher={Nature Publishing Group UK London},
  doi={https://www.nature.com/articles/s41535-024-00684-w}
}

@article{aggarwal2016unconventional,
  title={Unconventional superconductivity at mesoscopic point contacts on the 3D Dirac semimetal Cd$_3$As$_2$},
  author={Aggarwal, Leena and Gaurav, Abhishek and Thakur, Gohil S and Haque, Zeba and Ganguli, Ashok K and Sheet, Goutam},
  journal={Nat. Mater.},
  volume={15},
  number={1},
  pages={32--37},
  year={2016},
  publisher={Nature Publishing Group UK London},
  doi={https://www.nature.com/articles/nmat4455}
}

@article{he2016pressure,
  title={Pressure-induced superconductivity in the three-dimensional topological Dirac semimetal Cd$_3$As$_2$},
  author={He, Lanpo and Jia, Yating and Zhang, Sijia and Hong, Xiaochen and Jin, Changqing and Li, Shiyan},
  journal={npj Quantum Mater.},
  volume={1},
  number={1},
  pages={1--5},
  year={2016},
  publisher={Nature Publishing Group},
  doi={https://www.nature.com/articles/npjquantmats201614}
}

@article{wang2016observation,
  title={Observation of superconductivity induced by a point contact on 3D Dirac semimetal Cd$_3$As$_2$ crystals},
  author={Wang, He and Wang, Huichao and Liu, Haiwen and Lu, Hong and Yang, Wuhao and Jia, Shuang and Liu, Xiong-Jun and Xie, XC and Wei, Jian and Wang, Jian},
  journal={Nat. Mater.},
  volume={15},
  number={1},
  pages={38--42},
  year={2016},
  publisher={Nature Publishing Group UK London},
  doi={https://www.nature.com/articles/nmat4456}
}

@article{suslov2019observation,
  title={Observation of subkelvin superconductivity in Cd$_3$As$_2$ thin films},
  author={Suslov, AV and Davydov, AB and Oveshnikov, LN and Morgun, LA and Kugel, KI and Zakhvalinskii, VS and Pilyuk, EA and Kochura, AV and Kuzmenko, AP and Pudalov, VM and Aronzon, BA},
  journal={Phys. Rev. B},
  volume={99},
  number={9},
  pages={094512},
  year={2019},
  publisher={APS},
  doi={https://journals.aps.org/prb/abstract/10.1103/PhysRevB.99.094512}
}

@article{tsipas2018massless,
  title={Massless {D}irac fermions in ZrTe$_2$ semimetal grown on InAs (111) by van der Waals epitaxy},
  author={Tsipas, Polychronis and Tsoutsou, Dimitra and Fragkos, Sotirios and Sant, Roberto and Alvarez, Carlos and Okuno, Hanako and Renaud, Gilles and Alcotte, Reynald and Baron, Thierry and Dimoulas, Athanasios},
  journal={ACS {N}ano},
  volume={12},
  number={2},
  pages={1696--1703},
  year={2018},
  publisher={ACS Publications},
  doi={https://pubs.acs.org/doi/full/10.1021/acsnano.7b08350}
}

@article{muhammad2019transition,
  title={Transition from semimetal to semiconductor in ZrTe$_2$ induced by Se substitution},
  author={Muhammad, Zahir and Zhang, Bo and Lv, Haifeng and Shan, Huan and Rehman, Zia Ur and Chen, Shuangming and Sun, Zhe and Wu, Xiaojun and Zhao, Aidi and Song, Li},
  journal={ACS {N}ano},
  volume={14},
  number={1},
  pages={835--841},
  year={2019},
  publisher={ACS Publications},
  doi={https://pubs.acs.org/doi/full/10.1021/acsnano.9b07931}
}

@article{Wu_NatMat,
  title={The field-free {Josephson} diode in a van der {Waals} heterostructure},
  author={Wu, Heng  and Wang, Yaojia  and Xu, Yuanfeng  and Sivakumar, Pranava K. and Pasco, Chris and Filippozzi,  Ulderico and Parkin, Stuart S. P. and Zeng, Yu-Jia and McQueen, Tyrel and Ali, Mazhar N.},
  journal={Nat. Mater.},
  volume={604},
  number={8},
  pages={653--656},
  year={2022},
  publisher={Nature Publishing Group UK London},
  doi={https://www.nature.com/articles/s41586-022-04504-8}
}

@article{
Davydova_SciAd,
author = {Margarita Davydova  and Saranesh Prembabu  and Liang Fu },
title = {Universal {Josephson} diode effect},
journal = {Sci. Adv.},
volume = {8},
number = {23},
pages = {eabo0309},
year = {2022},
doi = {https://www.science.org/doi/abs/10.1126/sciadv.abo0309}
}

@misc{Islam_Supp,
  author       = "",
  title        = "See supplemental material for further details about sample growth, material characterization.",
  howpublished = "",
  month        = "",
  year         = "2025",
  note         = ""
}

@article{Yasuda_NComm,
	author = {Yasuda, Kenji and Yasuda, Hironori and Liang, Tian and Yoshimi, Ryutaro and Tsukazaki, Atsushi and Takahashi, Kei S. and Nagaosa, Naoto and Kawasaki, Masashi and Tokura, Yoshinori},
	date = {2019/06/21},
	doi = {10.1038/s41467-019-10658-3},
	id = {Yasuda2019},
	isbn = {2041-1723},
	journal = {Nat. Commun.},
	number = {1},
	pages = {2734},
	title = {Nonreciprocal charge transport at topological insulator/superconductor interface},
	doi = {https://doi.org/10.1038/s41467-019-10658-3},
	volume = {10},
	year = {2019},
	bdsk-doi-1 = {https://doi.org/10.1038/s41467-019-10658-3}}

@article{koteski2017first,
  title={First-principles calculations of tetragonal FeX (X= S, Se, Te): Magnetism, hyperfine-interaction, and bonding},
  author={Koteski, V and Ivanovski, Valentin N and Umi{\'c}evi{\'c}, Ana and Belo{\v{s}}evi{\'c}-{\v{C}}avor, Jelena and Toprek, Dragan and Mahnke, H-E},
  journal={J. Magn. Magn. Mater},
  volume={441},
  pages={769--775},
  year={2017},
  publisher={Elsevier},
  doi={https://www.sciencedirect.com/science/article/abs/pii/S0304885317308466}
}

@article{ma2025superconducting,
  title={Superconducting Diode Effects: Mechanisms, Materials and Applications},
  author={Ma, Jiajun and Zhan, Ruiya and Lin, Xiao},
  journal={Adv. Physics Res},
  pages={2400180},
  year={2025},
  publisher={Wiley Online Library},
  doi={https://advanced.onlinelibrary.wiley.com/doi/pdf/10.1002/apxr.202400180}
}

@article{Yan_2025,
title={Meissner Effect and Nonreciprocal Charge Transport in Superconducting {1T-CrTe$_2$/FeTe} Heterostructures},
author = {Yan, Zi-Jie and Chan, Ying-Ting and Yuan,Wei and Wang, Annie G. and Yi, Hemian and Wang, Zihao and Zhou, Lingjie and Rong, Hongtao and Zhuo, Deyi and Wang, Ke and Singleton, John and Winter, Laurel E. and Wu, Weida and Chang, Cui-Zu },
journal={arXiv:2412.09354},
year={2024},
doi={https://arxiv.org/abs/2412.09354}
}

@article{liu2009charge,
  title={Charge-carrier localization induced by excess Fe in the superconductor {Fe$_{1+y}$Te$_{1-x}$Se$_x$}},
  author={Liu, TJ and Ke, X and Qian, B and Hu, J and Fobes, D and Vehstedt, EK and Pham, H and Yang, JH and Fang, MH and Spinu, L and Schiffer, P and Liu. Y and Mao, ZQ},
  journal={Phys. Rev. B},
  volume={80},
  number={17},
  pages={174509},
  year={2009},
  publisher={APS},
  doi={https://journals.aps.org/prb/abstract/10.1103/PhysRevB.80.174509}
}

@article{lee2019spin,
  title={Spin scattering and noncollinear spin structure-induced intrinsic anomalous Hall effect in antiferromagnetic topological insulator MnBi$_2$Te$_4$},
  author={Lee, Seng Huat and Zhu, Yanglin and Wang, Yu and Miao, Leixin and Pillsbury, Timothy and Yi, Hemian and Kempinger, Susan and Hu, Jin and Heikes, Colin A and Quarterman, Patrick and others},
  journal={Phys. Rev. Research},
  volume={1},
  number={1},
  pages={012011},
  year={2019},
  publisher={APS},
  doi={https://journals.aps.org/prresearch/abstract/10.1103/PhysRevResearch.1.012011}
}

@article{chu2023broad,
  title={Broad and colossal edge supercurrent in Dirac semimetal Cd$_3$As$_2$ Josephson junctions},
  author={Chu, Chun-Guang and Chen, Jing-Jing and Wang, An-Qi and Tan, Zhen-Bing and Li, Cai-Zhen and Li, Chuan and Brinkman, Alexander and Xiang, Peng-Zhan and Li, Na and Pan, Zhen and Lu, Hai-Zhou and Yu, Dapeng and Liao, Zhi-Min},
  journal={Nat. Commun.},
  volume={14},
  number={1},
  pages={6162},
  year={2023},
  doi={https://www.nature.com/articles/s41467-023-41815-4#Sec7},
  publisher={Nature Publishing Group UK London}
}

@article{li2020fermi,
  title={Fermi-arc supercurrent oscillations in Dirac semimetal Josephson junctions},
  author={Li, Cai-Zhen and Wang, An-Qi and Li, Chuan and Zheng, Wen-Zhuang and Brinkman, Alexander and Yu, Da-Peng and Liao, Zhi-Min},
  journal={Nat. Commun.},
  volume={11},
  number={1},
  pages={1150},
  year={2020},
 doi={https://www.nature.com/articles/s41467-020-15010-8#Sec7},
publisher={Nature Publishing Group UK London}
}

@book{ichimiya2004reflection,
  title={Reflection high-energy electron diffraction},
  author={Ichimiya, Ayahiko and Cohen, Philip I},
  year={2004},
  publisher={Cambridge University Press},
  doi={https://www.cambridge.org/core/books/reflection-highenergy-electron-diffraction/162FE7186C89C6A8269619B7EDF5F1E8}
}

@book{hasegawa2012reflection,
  title={Reflection high-energy electron diffraction},
  author={Hasegawa, Shuji},
  journal={Characterization of Materials},
  volume={97},
  pages={1925--1938},
  year={2012},
  publisher={John Wiley \& Sons Hoboken, NJ, USA},
  doi={https://onlinelibrary.wiley.com/doi/book/10.1002/0471266965}
}

@article{park1995resistance,
  title={Resistance anomaly and excess voltage near superconducting interfaces},
  author={Park, M and Isaacson, MS and Parpia, JM},
  journal={Phys. Rev. Lett.},
  volume={75},
  number={20},
  pages={3740},
  year={1995},
  publisher={APS},
  doi={https://journals.aps.org/prl/abstract/10.1103/PhysRevLett.75.3740}
}

@article{zhang2013metal,
  title={Metal--bosonic insulator--superconductor transition in boron-doped granular diamond},
  author={Zhang, Gufei and Zeleznik, Monika and Vanacken, Johan and May, Paul W and Moshchalkov, Victor V},
  journal={Phys. Rev. Lett.},
  volume={110},
  number={7},
  pages={077001},
  year={2013},
  publisher={APS},
  doi={https://journals.aps.org/prl/abstract/10.1103/PhysRevLett.110.077001}
}

@article{si2010superconductivity,
  title={Superconductivity in epitaxial thin films of Fe$_{1.08}$Te:O$_x$},
  author={Si, Weidong and Jie, Qing and Wu, Lijun and Zhou, Juan and Gu, Genda and Johnson, PD and Li, Qiang},
  journal={Phys. Rev. B.},
  volume={81},
  number={9},
  pages={092506},
  year={2010},
  publisher={APS},
  doi={https://journals.aps.org/prb/abstract/10.1103/PhysRevB.81.092506}
}

@article{cao2018unconventional,
  title={Unconventional superconductivity in magic-angle graphene superlattices},
  author={Cao, Yuan and Fatemi, Valla and Fang, Shiang and Watanabe, Kenji and Taniguchi, Takashi and Kaxiras, Efthimios and Jarillo-Herrero, Pablo},
  journal={Nature},
  volume={556},
  number={7699},
  pages={43--50},
  year={2018},
  publisher={Nature Publishing Group UK London},
  doi={https://www.nature.com/articles/nature26160}
}

@article{park2021tunable,
  title={Tunable strongly coupled superconductivity in magic-angle twisted trilayer graphene},
  author={Park, Jeong Min and Cao, Yuan and Watanabe, Kenji and Taniguchi, Takashi and Jarillo-Herrero, Pablo},
  journal={Nature},
  volume={590},
  number={7845},
  pages={249--255},
  year={2021},
  publisher={Nature Publishing Group UK London},
  doi={https://www.nature.com/articles/s41586-021-03192-0#additional-information}
}

@article{balakrishnan2025complex,
  title={Complex Magnetic Ordering in Candidate Topological Superconductors},
  author={Balakrishnan, Purnima P and Yi, Hemian and Yan, Zi-Jie and Yuan, Wei and Suter, Andreas and Jensen, Christopher J and Manuel, Pascal and Orlandi, Fabio and Hanashima, Takayasu and Kinane, Christy J and Caruana, AJ and Maranville, BB and Salman,Z and Prokscha,T and Chang, Cui-Zu and Grutter, AJ},
  journal={arXiv:2503.11502},
  year={2025},
  doi={https://arxiv.org/pdf/2503.11502}
}

@article{venditti2019nonlinear,
  title={Nonlinear I-V characteristics of two-dimensional superconductors: Berezinskii-Kosterlitz-Thouless physics versus inhomogeneity},
  author={Venditti, G and Biscaras, J and Hurand, S and Bergeal, N and Lesueur, J and Dogra, A and Budhani, RC and Mondal, Mintu and Jesudasan, John and Raychaudhuri, Pratap and Caprara, S and Benfatto, L},
  journal={Phys. Rev. B},
  volume={100},
  number={6},
  pages={064506},
  year={2019},
  publisher={APS},
  doi={https://journals.aps.org/prb/abstract/10.1103/PhysRevB.100.064506}
}

@article{kumar2023flux,
  title={Flux-flow instability across Berezinskii Kosterlitz Thouless phase transition in KTaO$_3$ (111) based superconductor},
  author={Kumar Ojha, Shashank and Mandal, Prithwijit and Kumar, Siddharth and Maity, Jyotirmay and Middey, Srimanta},
  journal={Commun. Phys.},
  volume={6},
  number={1},
  pages={126},
  year={2023},
  publisher={Nature Publishing Group UK London},
  doi={https://www.nature.com/articles/s42005-023-01251-8}
}

@article{yan2026stoichiometric,
  title={Stoichiometric {FeTe} is a superconductor},
  author={Yan, Zi-Jie and Wang, Zihao and Xia, Bing and Paolini, Stephen and Chan, Ying-Ting and Dihingia, Nikalabh and Rong, Hongtao and Xiao, Pu and Halanayake, Kalana D and Song, Jiatao and others},
  journal={Nature},
  pages={1--7},
  year={2026},
  publisher={Nature Publishing Group UK London},
  doi={https://www.nature.com/articles/s41586-026-10321-0}
}

@article{nadeem2023superconducting,
  title={The superconducting diode effect},
  author={Nadeem, Muhammad and Fuhrer, Michael S and Wang, Xiaolin},
  journal={Nature Reviews Physics},
  volume={5},
  number={10},
  pages={558--577},
  year={2023},
  publisher={Nature Publishing Group UK London},
  doi={https://www.nature.com/articles/s42254-023-00632-w}
}

@article{yuan2022supercurrent,
  title={Supercurrent diode effect and finite-momentum superconductors},
  author={Yuan, Noah FQ and Fu, Liang},
  journal={Proc. Nat. Acad. Sci.},
  volume={119},
  number={15},
  pages={e2119548119},
  year={2022},
  publisher={National Academy of Sciences},
  doi={https://www.pnas.org/doi/10.1073/pnas.2119548119}
}

\newpage
    
\begin{figure}
       \centering
        \includegraphics[width=16cm]{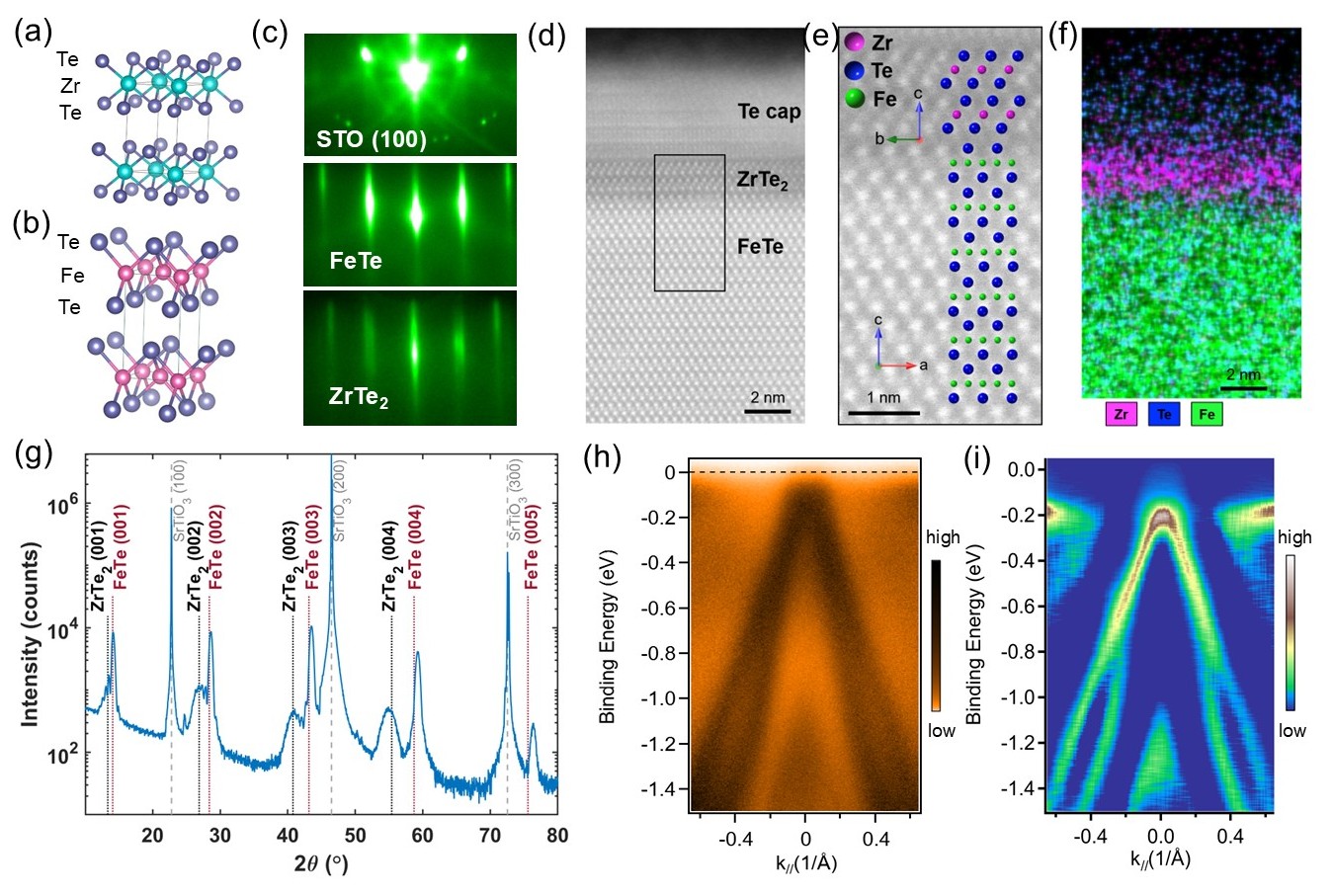}
        \caption{(a) Crystal structure of ZrTe$_2$. (b) Crystal structure of FeTe. (c) RHEED images captured during different stages of growth: the pre-growth STO substrate (top panel), after  $20$ nm growth of FeTe (middle), and after $3$~nm growth of \ZrTe~(bottom). (d) Cross-sectional HAADF-STEM image showing the atomically sharp interface between ZrTe$_2$ and FeTe. (e) Zoomed-in region of the HAADF-STEM image in (d) (the area is marked by the black box). (f) An EDX map from a similar region to that shown in (d). (g) XRD $2 \theta-\omega$ scans of a \ZrTe~(6 UC)/FeTe (35 UC)/SrTiO$_3$(100) heterostructure, capped with Te. (h) Angle-resolved photo-emission spectra (ARPES) of a 4UC ZrTe$_2$/FeTe heterostructure, taken at $T = 300$ K along the $\bar{M} - \bar{\Gamma}- \bar{M}$ direction using 21.2 eV excitation from a helium lamp. (i) The second derivative of the spectra displayed in (h).} 
        \label{1}
\end{figure}
    
\begin{figure}
       \centering
        \includegraphics[width=12cm]{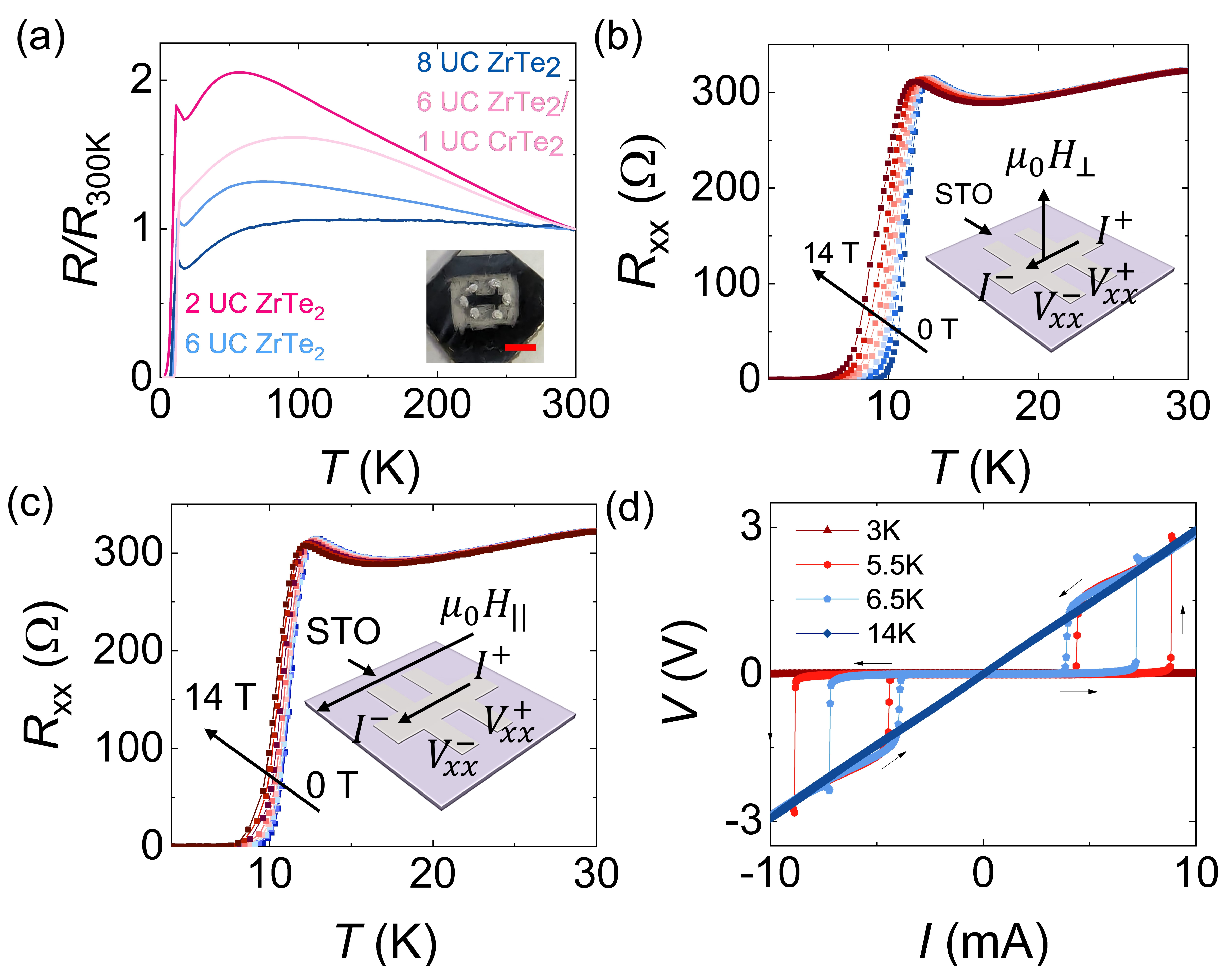}
        \caption{Electrical transport properties: (a) Longitudinal resistance normalized by $300$~K magnitude ($R/R_{300K}$) vs temperature ($T$) for different films. Inset shows optical image of a scratched Hall bar device. The scale bar is $1$~mm. (b) $R_{xx}$ vs $T$ as a function of magnetic field $\mu_0H$ directed perpendicular to the sample plane. (c)  Similar data as (b) with magnetic field in plane and parallel to the current. (d) Current-voltage characteristic vs. temperature, showing clear switching and hysteresis.} 
        \label{2}
\end{figure}
    
\begin{figure}
       \centering
        \includegraphics[width=12cm]{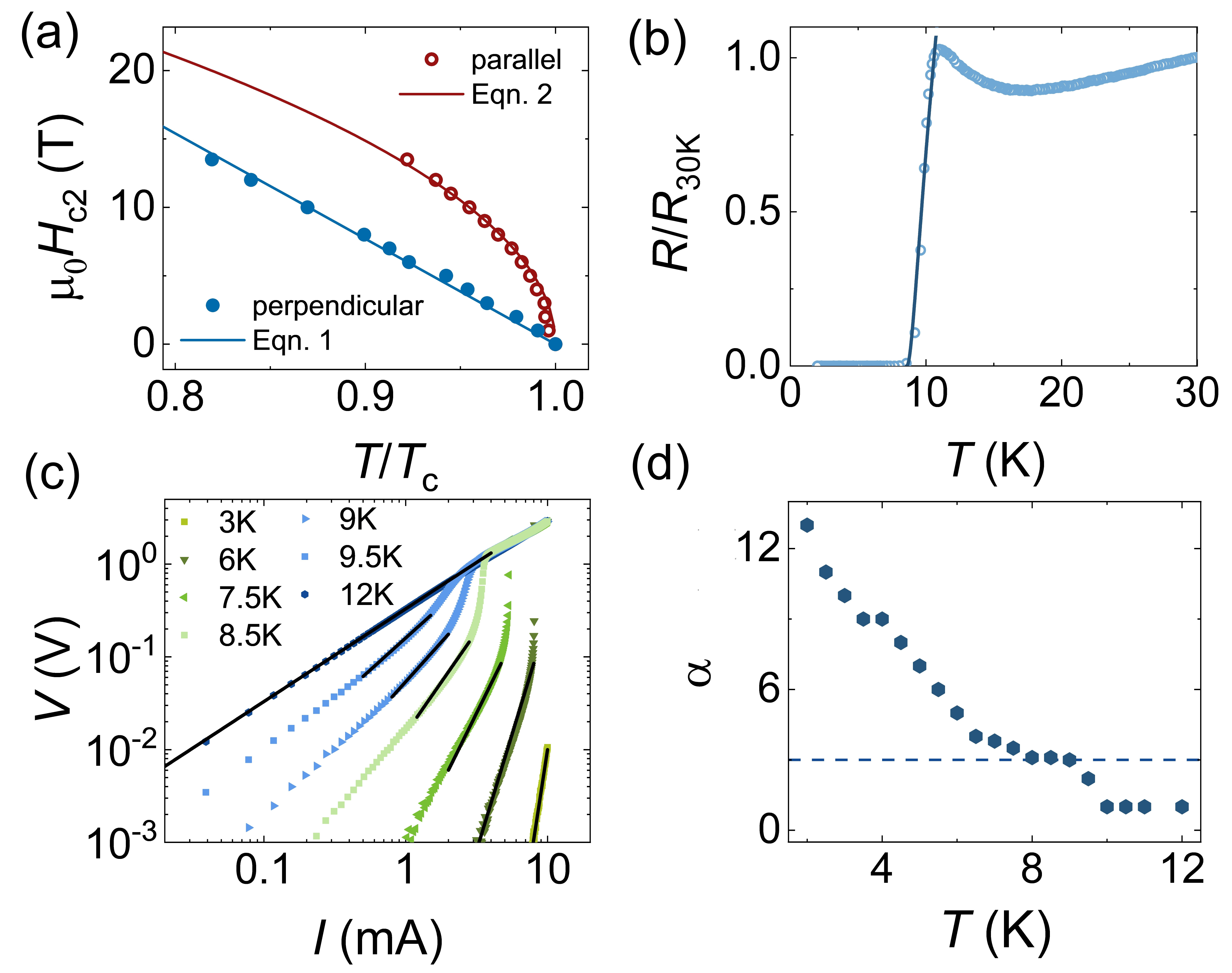}
        \caption{Two-dimensional superconductivity: (a) Upper critical magnetic field ($H_{c2}$) as a function of reduced temperature ($T/T_c$). The solid black and blue lines are fits to the data according to Eqns. 1 and 2. (b) $R/R_{30K}$ vs. $T$ for the 6 UC sample. The solid line is fit to the data according to the Halperin-Nelson equation. (c) Current-voltage characteristic plotted on a log-log scale. Solid lines are fits to the data according to $V\propto I^\alpha$. (d) $\alpha$ as a function of temperature, showing a rapid increase below the transition.} 
        \label{3}
\end{figure}
    
    \begin{figure}
        \centering
        \includegraphics[width=12cm]{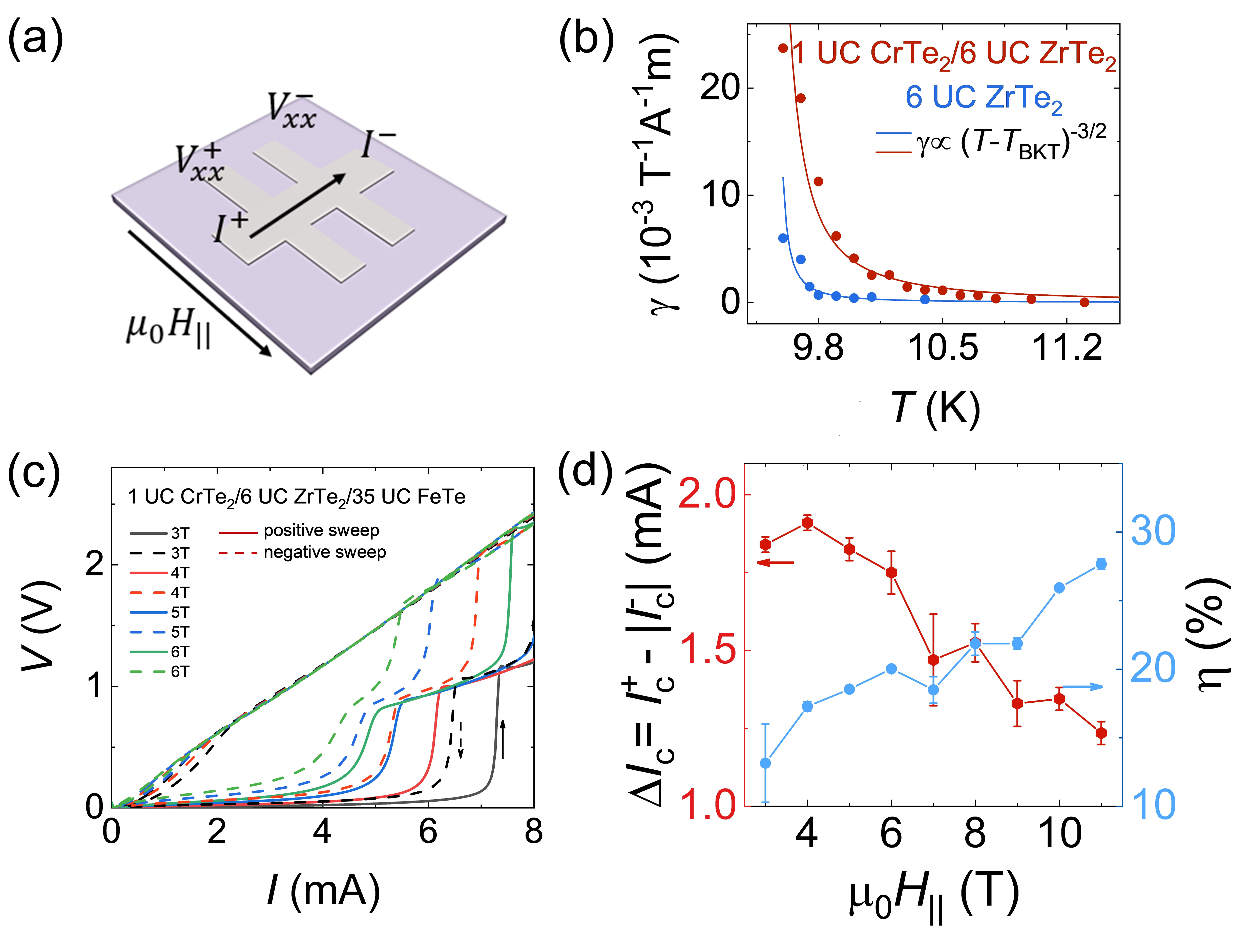}
        \caption{Non-reciprocal charge transport: (a) Schematic of the second harmonic resistance measurement to extract magneto-chiral anisotropy: the current ($I$) and magnetic field ($B$) are in-plane and perpendicular to each other. (b) The temperature dependence of the magneto-chiral coefficient $\gamma$ measured at $I=600$~$\mu$A. The solid line is fit to the data using $\gamma~\propto (T-T_{\rm{BKT}})^{-3/2}$. (c) Current-voltage characteristics for 1UC CrTe$_2$/6UC ZrTe$_2$/35UC  at representative magnetic fields $3$~T, $4$~T, $5$~T, $6$~T, showing asymmetry between the critical current between the positive and negative sweep directions. The solid (dashed) lines represent positive (negative) sweep direction. (d) Left axis: the difference between the magnitude of the critical current between positive and negative current sweep directions, $\Delta I_c= I_{c}^{+}-|I^-_c|$ plotted as a function of magnetic field. Right axis: the efficiency $\eta = \frac{I^+_c-|I^-_c|}{I^+_c+|I^-_c|}$ is plotted as a function of magnetic field.} 
        \label{4}
    \end{figure}
    
    \end{document}